\documentclass[english,english,letterpaper,prl,aps,superscriptaddress,floatfix,twocolumn,showpacs]{revtex4}
\usepackage[T1]{fontenc}
\usepackage[latin9]{inputenc}
\usepackage{amssymb}
\usepackage{graphicx}

\makeatletter

\@ifundefined{textcolor}{}
{%
 \definecolor{BLACK}{gray}{0}
 \definecolor{WHITE}{gray}{1}
 \definecolor{RED}{rgb}{1,0,0}
 \definecolor{GREEN}{rgb}{0,1,0}
 \definecolor{BLUE}{rgb}{0,0,1}
 \definecolor{CYAN}{cmyk}{1,0,0,0}
 \definecolor{MAGENTA}{cmyk}{0,1,0,0}
 \definecolor{YELLOW}{cmyk}{0,0,1,0}
}

\makeatother

\usepackage{babel}
\begin{document}

\title{Proposal for an Optomechanical Microwave Sensor at the Subphoton level}

\author{Keye Zhang}
\affiliation{Quantum Institute for Light and Atoms, State Key Laboratory of Precision Spectroscopy, Department of Physics, East
China Normal University, Shanghai, 200241, China  }
\affiliation{B2 Institute, Department of Physics and College of Optical Sciences,
University of Arizona, Tucson, Arizona 85721, USA}
\author{Francesco Bariani}
\affiliation{B2 Institute, Department of Physics and College of Optical Sciences,
University of Arizona, Tucson, Arizona 85721, USA}
\author{Ying Dong}
\affiliation{B2 Institute, Department of Physics and College of Optical Sciences,
University of Arizona, Tucson, Arizona 85721, USA}
\affiliation{Department of Physics, Hangzhou Normal University, Hangzhou, Zhejiang 310036, China}
\author{Weiping Zhang}
\affiliation{Quantum Institute for Light and Atoms, State Key Laboratory of Precision Spectroscopy, Department of Physics, East
China Normal University, Shanghai, 200241, China  }
\author{Pierre Meystre}
\affiliation{B2 Institute, Department of Physics and College of Optical Sciences,
University of Arizona, Tucson, Arizona 85721, USA}

\begin{abstract}
Because of their low energy content, microwave signals at the single-photon level are extremely challenging to measure. Guided by recent progress in single-photon optomechanics and hybrid optomechanical systems, we propose a multimode optomechanical transducer that can detect intensities significantly below the single-photon level via adiabatic transfer of the microwave signal to the optical frequency domain where the measurement is then performed. The influence of intrinsic quantum and thermal fluctuations is also discussed.
\end{abstract}
\maketitle

\emph{Introduction}.---The microwave frequency domain of the electromagnetic spectrum is the stage of a wealth of phenomena, ranging from the determination of the quantum energy levels of superconductor nanostructures to the rotational modes of molecules and to the characterization of the cosmic microwave background.  Several detection schemes sensitive to microwave radiation at the single-photon level have been demonstrated. Examples include semiconductor quantum dots in high magnetic field~\cite{Komiyama2000}, circular Rydberg atoms in cavity QED setups~\cite{Raimond2001, Guerlin2007, Peaudecerf2014}, and superconducting qubits in circuit QED~\cite{Houck2007, Johnson2010}.  An alternative approach involves the use of linear amplifiers~\cite{ClerkRMP}. These devices allow the reconstruction of average amplitudes~\cite{Houck2007} and correlation functions~\cite{Bozyigit2011} and may operate both as phase-preserving (insensitive)~\cite{Devoret} and phase-sensitive~\cite{Yurke1988,Castellanos2008} amplifiers, but they require an integration over many events to achieve a sizable signal.

Even though there have been proposals and experiments to realize a photon multiplier in the microwave regime \cite{Romero2009,Chen2011,Govia2014}, no general purpose efficient single-photon detector has been developed so far, as photon energies in that frequency domain are in the milli-electron volt range, 3 orders of magnitude smaller than in the visible or near-infrared spectral regions. On the other hand, in the optical frequency domain a variety of ultra-sensitive detectors have been developed over the past sixty years. This suggests that an alternative route for the detection of feeble microwave signals is via their conversion to the optical frequency domain. Photonic front-end microwave receivers based on the electro-optical effect~\cite{Ilchenko2002} and atomic interfaces based on electromagnetically induced transparency have exploited nonlinear conversion to this end~\cite{Hafezi2012, Sedlacek2012}. The main limitations in sensitivity are the small strength of the interaction and the fluctuations of the optical driving fields.

Recent advances in nano- and optomechanics offer an attractive approach to engineer interactions of light and mechanics that achieve that goal via the radiation pressure force; see Ref.~\cite{OMReview} for recent reviews.  Several theoretical proposals have considered the optomechanically mediated quantum state transfer between microwave and optical fields \cite{Barzanjeh2012, Wang2012, Tian2012, Zhangqi2014} and have emphasized the potential of hybrid  systems as quantum information interfaces~\cite{Stannigel2010, Tsang2010, Regal2011, Barzanjeh2011}, in which case state transfer fidelity is of particular interest. Developments of particular relevance include the experimental realization of coherent conversion between microwave and optical field based on a hybrid optomechanical setup~\cite{Andrews2014, Bochmann2013, Bagci2014}. The present work has the different goal to convert the {\em mean intensity} of a feeble, narrow band microwave signal to a signal  at an optical frequency where detection can proceed by traditional methods. 

One key aspect of this proposed detector is that it relies on an off-resonant, multimode process. This is motivated by the need to manage and minimize the thermal mechanical noise, as well as to circumvent the effect of the fluctuations of the driving electromagnetic fields required to ensure a strong enough optomechanical coupling. These sources of noise can be significantly reduced by (i) working in a far off-resonant regime with respect to the mechanics, (ii) using pumping fields that drive ancillary cavity modes different from those at the signal frequencies, for both microwave and optical, and (iii) exploiting the polariton modes of the cavity-mechanics system to perform the frequency conversion of the signal  via a modulation of the detuning of the optical pump. 

\begin{figure}[ptbh]
\includegraphics[width=3.5in]{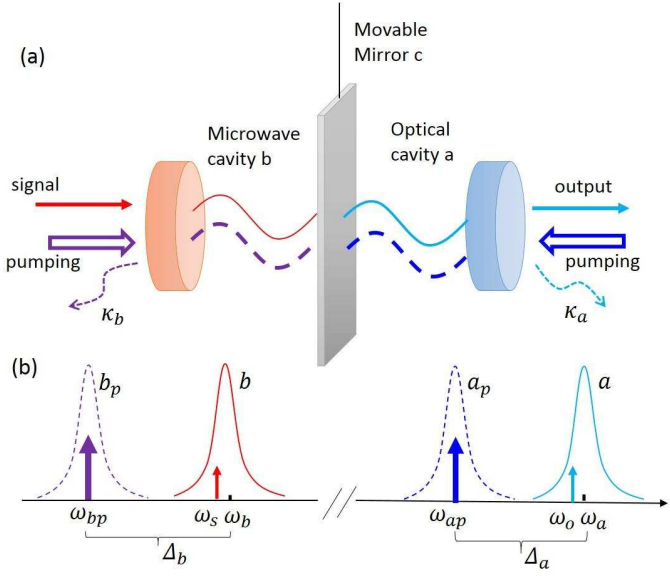}
\caption{(a) Dual-cavity optomechanical system. (b) Sketch of the heterodynelike pumping scheme with the microwave signal and the driving field near resonant with cavity mode $b$ and ancilliary mode $b_p$, respectively. Similarly in the optical side.}
\label{fig:scheme}
\end{figure}

\emph{The system}.--- The proposed sensor is composed of a mechanical oscillator optomechanically coupled to both a microwave and an optical multimode resonator; see idealized setup in Fig.~\ref{fig:scheme}.  

Consider first the microwave cavity. To avoid the noise connected with the pumping field while still maintaining a large optomechanical coupling strength, we adopt a multimode configuration where a strong optomechanical coupling is provided by an auxiliary field at frequency $\omega_{bp}$ different from that of the signal to be detected, see Fig.~\ref{fig:scheme}(b). This three-mode optomechanical interaction is described by the Hamiltonian~\cite{Miao2009, Cheung2011, Mari2012}
\begin{equation}
V_{3m}=\hbar g_{b0}(\hat{b}_p+ \hat{b})^{\dagger}(\hat{b}_p+\hat{b})(\hat{c}+\hat{c}^{\dagger}),
\label{hamilt 1}
\end{equation}
where $g_{b0}$ is the single microwave photon optomechanical coupling constant.
We assume that $\omega_{bp}$ is resonant with a longitudinal cavity mode, while the signal field $\hat b$, assumed to be extremely weak, is slightly detuned from another mode of frequency $\omega_b$. In the displaced picture for $\hat b_p$ and $\hat{c}$, $\hat b_p \rightarrow \beta_p + \hat b_p$ and $\hat{c} \rightarrow \mathcal{C} + \hat{c}$, the Hamiltonian~(\ref{hamilt 1}) becomes
\begin{eqnarray}
V_{3m, \rm eff}&=&\hbar G_b(\hat b+\hat b^{\dagger})(\hat{c}+\hat{c}^{\dagger})+\hbar x_{c}g_{b0}(\hat{b}_p\hat{b}^{\dagger}+\hat{b}_p^{\dagger}\hat b)\nonumber \\
&+&\hbar G_b(\hat{b}_p+\hat{b}_p^{\dagger})(\hat{c}+\hat{c}^{\dagger}).
\label{hamilt 2}
\end{eqnarray}
The first term is the usual linearized optomechanical coupling between the signal mode $\hat{b}$ and phonon mode $\hat{c}$ with strength $G_b=\beta_p g_{b0}$. We assume that the pump field is phase locked so that $G_b$ is real and positive. Its fluctuations feed into the system as noise through the second and the third terms of $V_{3m, \rm eff}$ which arise from the scattering and the optomechanical coupling of the pumped mode, respectively. The second term, proportional to the steady position quadrature  of the phonon field, $x_c=\mathcal{C} +\mathcal{C}^{*}$, can be safely neglected under the condition $|x_c| \ll |\beta_p|$ which is easily realized \cite{Supple} in the mirror-in-the-middle geometry of Fig.~\ref{fig:scheme}. Finally, the third term results in contributions to the system dynamics at a frequency that differs from the first term by $\pm (\omega_b-\omega_{bp})$.  This difference is of the order of the free spectral range of the cavity (for longitudinal modes) so that it can easily be filtered out in a manner familiar from heterodyne detection. For the narrow band detection scheme considered here it is therefore sufficient to keep only the first term in the Hamiltonian (\ref{hamilt 2}).

Following a similar argument for the optical fields, the effective Hamiltonian for the full system becomes
\begin{eqnarray}
H & = & \hbar\omega_{m}\hat{c}^{\dagger}\hat{c}-\hbar\Delta_{a}\hat{a}^{\dagger}\hat{a}-\hbar\Delta_{b}\hat{b}^{\dagger}\hat{b}\nonumber \\
 & + &\hbar G_{a}(\hat{a}+\hat{a}^{\dagger})(\hat{c}+\hat{c}^{\dagger})+\hbar G_{b}(\hat{b}+\hat{b}^{\dagger})(\hat{c}+\hat{c}^{\dagger}),\label{eq:original H}
\end{eqnarray}
where $\hat{a}$, $\hat{b}$, and $\hat{c}$ are the (displaced) annihilation operators for the optical, microwave, and mechanical modes with corresponding frequencies $\omega_{a}$, $\omega_{b}$, and $\omega_{m}$. The optical and microwave cavity-pump detunings are $\Delta_{a}=\omega_{ap}-\omega_{a}+x_{c}g_{a0}$  and $\Delta_{b}=\omega_{bp}-\omega_{b}+x_{c}g_{b0}$, respectively, with $\omega_{ap}$ and $\omega_{bp}$ the frequencies of the optical and microwave pumps. $G_a$ and $G_b$ are the effective optomechanical coupling strength set by the steady amplitude of the pumped ancillary optical and microwave cavity modes. Note that $G_{a,b}$ are of opposite signs and the equilibrium position of the mechanical resonator is set by the relative strength of the two pumps, so that the microwave drive needs to have a significantly stronger light flux than the optical pump.

In the resonant situation $\Delta_{a}=\Delta_{b}=-\omega_{m}$, an effective interaction follows from performing the rotating wave approximation, which gives $H_{I}=\hbar G_{a}(\hat{a}\hat{c}^{\dagger}+\hat{c}\hat{a}^{\dagger})+\hbar G_{b}(\hat{b}\hat{c}^{\dagger}+\hat{c}\hat{b}^{\dagger})$. If $G_{a}$ and $G_{b}$ are appropriately modulated in time the system then adiabatically follows a superposition of cavity modes $\hat{a}$ and $\hat{b}$ without any population of the mechanical mode $\hat{c}$ (dark mode) \cite{Wang2012, Tian2012}. In contrast, for the off-resonant case considered here, $\Delta_{a,b}\neq\omega_m$, the microwave and optical fields are coupled by a three-level Raman-like interaction via the mechanical mode. 

\begin{figure}
\includegraphics[width=3.5in]{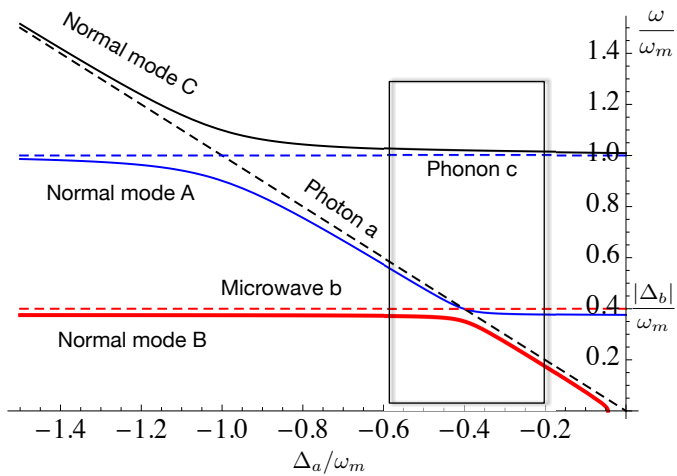}
\caption{Eigenfrequencies of the normal modes (polaritons) as functions of optical detuning $\Delta_{a}/\omega_{m}$ for the case $-G_{a}/\omega_{m}=G_{b}/\omega_{m}=0.1$ and $\Delta_{b}/\omega_{m}=-0.4$. Dashed lines: noninteracting energies of the bare modes. We have framed the part of the spectrum spanned by $\Delta_{a}$ during the conversion process.\label{fig:normal spectrum}}
\end{figure}

\emph{Normal mode picture}.---To discuss the microwave-to-optical conversion process in this effective three-mode configuration, it is convenient to switch to a normal mode (polariton) representation of the system \cite{ourPRL}. After removing a constant term, the Hamiltonian (\ref{eq:original H}) can be recast in the diagonal form $H=\hbar\omega_{A}\hat{A}^{\dagger}\hat{A}+\hbar\omega_{B}\hat{B}^{\dagger}\hat{B}+\hbar\omega_{C}\hat{C}^{\dagger}\hat{C}$,  where $\hat{A}$, $\hat{B},$ and $\hat{C}$ are the boson annihilation operators for the normal mode excitations. In general, these are superpositions of the optical, microwave, and mechanical modes.  Figure~\ref{fig:normal spectrum} shows their frequencies $\omega_{A,B,C}$ as functions of the optical detuning $\Delta_{a}$.  At the mechanical resonance, $\Delta_{a}=-\omega_{m}$, the degeneracy between the optical photon and the phonon is lifted by the optomechanical interaction, with an energy splitting of the order of $2G_{a}$. A second avoided crossing occurs at the resonance between optical and microwave photons, $\Delta_{a}=\Delta_{b}$,  with a splitting of the order of $4G_{a}G_{b}/\omega_{m}$ resulting from the indirect coupling between the electromagnetic modes via the mechanical mode. 

We focus on the region close to the microwave-optical resonance framed in Fig.~\ref{fig:normal spectrum}. On the left-hand side, $\Delta_{a}<\Delta_{b}$ and $|\Delta_{a}-\Delta_{b}|\gg4|G_{a}G_{b}|/\omega_{m}$ the polariton $\hat{B}$ describes a microwavelike excitation, with $\omega_{B} \sim -\Delta_{b}$ and $\hat{B}\sim\hat{b}$, while for $\Delta_{a}>\Delta_{b}$, the polariton becomes optical-like  $\hat{B}\sim\hat{a}$ and annihilates an excitation of frequency $\omega_{B}\sim-\Delta_{a}$. The opposite holds for the polariton $A$, which is opticallike for $\Delta_{a}<\Delta_{b}$ and microwavelike on the other side of the resonance. The polariton $C$ remains phonon-like in this whole region, indicating that the dynamics of the mechanical excitation is decoupled from that of the electromagnetic fields. 

\emph{Conversion process}.---When $\Delta_{a}$ is slowly switched from the left-hand to the right-hand side of the resonance, the polariton $B$  adiabatically evolves from the microwavelike excitation to the optical-like excitation while conserving its population, $\langle\hat{B}^{\dagger}(t)\hat{B}(t)\rangle\approx\langle\hat{b}^{\dagger}(t_{0})\hat{b}(t_{0})\rangle$, where $\langle\hat{b}^{\dagger}(t_{0})\hat{b}(t_{0})\rangle$ accounts for both the input signal field to be measured and the microwave cavity noise. Likewise the polariton $A$, which is initially optical-like, evolves into a microwavelike excitation while maintaining its population $\langle\hat{A}^{\dagger}(t)\hat{A}(t)\rangle\approx\langle\hat{a}^{\dagger}(t_{0})\hat{a}(t_{0})\rangle=0$, where the last equality holds if the optical mode is initially in a vacuum, a condition easy to satisfy.

The adiabaticity of the transfer requires that  $\Delta_{a}$ be switched at a rate much slower than the interband separation, $1/\tau \ll 4|G_{a}G_{b}|/\omega_{m}$, where $\tau$ is the switching time. In addition it is also necessary that this operation occurs in a time short compared to the inverse decay rates of the polariton modes,  which are combinations of the cavity decay rates $\kappa_{a,b}$ and the mechanical damping rate $\gamma$. (This condition also ensures that $\alpha$ and $\beta$ remain constant during the switch of  $\Delta_{a}$. )

We describe the detection protocol as a time-gated three-step process. First, during a ``receiving'' time window $\tau_r$ that lasts until $t_0$, the optical detuning is fixed at $\Delta_{a}<\Delta_{b}$, with $|\Delta_{a}-\Delta_{b}|\gg4|G_{a}G_{b}|/\omega_{m}$ and the microwave cavity captures a narrow band signal that is stored in the mode $b$. During that time the optical mode  $a$ is in a vacuum and the microwave-optical field interaction is negligible due to their large mismatch in frequency. This is followed by a  ``transfer'' time interval $\tau$ starting at $t_0$ during which $\Delta_{a}$ is  switched to  $\Delta_{b}$ at a rate 
\begin{equation}
1/\kappa_{a,b}\gg \tau \gg\omega_{m}/4|G_{a}G_{b}|,
\label{time condition}
\end{equation}
resulting in the signal being transferred into an optical field without any significant coupling to the external reservoirs. Finally, during the detecting time window $\tau_d >t_{0}+\tau$ the interaction is quenched and the cavities couple with their environment, thus releasing the optical output field that can be measured by standard methods.

\emph{Input-output dynamics}.---The  analysis of the conversion of the microwave signal to the optical field can be performed in terms of Heisenberg-Langevin equations of motion $\partial_{t}\hat{u}=-i[\hat{u},\hat{H}]/\hbar-\kappa_{u}\hat{u}+\sqrt{2\kappa_{u}}\hat{u}_{\rm in}$, where $\hat{u}$ are the annihilation operators for the bare modes \{$\hat{a}, \hat{b}, \hat{c}$\}, $\kappa_u$ are their dissipation rates (with $\kappa_c \equiv \gamma$),  and $\hat{u}_{\rm in}$ account for the associated noise operators and input fields.  In the absence of input fields the nonvanishing noise correlations are $\langle\hat{u}_{\rm in}(t)\hat{u}_{\rm in}^{\dagger}(t^{\prime})\rangle=(\bar{n}_{u}+1)\delta(t-t^{\prime})$ and $\langle\hat{u}_{\rm in}^{\dagger}(t)\hat{u}_{\rm in}(t^{\prime})\rangle=\bar{n}_{u}\delta(t-t^{\prime})$, where $\bar{n}_{u}=1/[\exp(\hbar\omega_{u}/k_{B}T_{u})-1]$, $T_{u}$ being the temperature of the thermal reservoir of mode $u$. For the optical field $\bar{n}_{a}\approx0$ in practice.

In the far off-resonant case $\omega_{m}\gg |\Delta_{a,b}|,|G_{a,b}|,\kappa_{a,b}, \gamma$ we adiabatically eliminate the phonon mode $\hat{c}$ by inserting its formal solution $\hat{c}\approx[-G_{a}(\hat{a}+\hat{a}^{\dagger})-G_{b}(\hat{b}+\hat{b}^{\dagger})]/\omega_{m}$ into the equations for the modes $a$ and $b$ while retaining the mechanical noise term and neglecting the memory effect. The interaction between the microwave and optical modes is then described by the equation
\begin{equation}
\partial_{t}\hat{a} = (i\Delta_{a}^{\prime}-\kappa_a)\hat{a}+i\frac{2G_{a}^{2}}{\omega_{m}}\hat{a}^{\dagger}+iG^{\prime}(\hat{b}+\hat{b}^{\dagger})+\sqrt{2\kappa_{a}}\hat{a}_{\rm in}^{\prime},\label{eq:da}
\end{equation}
where $G^{\prime}=2G_{a}G_{b}/\omega_{m}$, and similarly for mode $b$ with $a\leftrightarrow b$ \cite{Supple}.

In the far off-resonant case, we must keep the antirotating terms in the optomechanical interaction when adiabatically eliminating the mechanics. This results in a squeezing contribution to the dynamics of $a$ and $b$ with the original detuning becoming  $\Delta_{a,b}^{\prime}=\Delta_{a,b}+2G_{a,b}^{2}/\omega_{m}$ and
\begin{equation}
\hat{a}_{\rm in}^{\prime}=\hat{a}_{\rm in} -iG_{a}\sqrt{\frac{\gamma}{\kappa_{a}}}\left[\int_{0}^{t}e^{(-i\omega_{m}-\gamma)(t-t^{\prime})}\hat{c}_{\rm in}(t^{\prime})dt^{\prime}+H.c.\right],
\label{eq:ain}
\end{equation}
and similarly for $b_{\rm in}^\prime$ with $a \rightarrow b$. When we focus on the signal fields of narrow linewidth around cavity modes, the noise autocorrelation functions approximately become $\langle\hat{a}_{\rm in}^{\prime}(t)\hat{a}_{\rm in}^{\prime\dagger}(t^{\prime})\rangle=(\bar{n}_{a}+m_{a}+1)\delta(t-t^{\prime})$ and $\langle\hat{a}_{\rm in}^{\prime}(t)\hat{a}_{\rm in}^{\prime}(t^{\prime})\rangle=-m_{a}\delta(t-t^{\prime})$, with $m_{a}=(G_{a}^{2}\gamma/\omega_{m}^{2}\kappa_{a})(2\bar{n}_{c}+1)$, with also the appearance of cross-correlations characteristic of a squeezed two-mode reservoir, $\langle\hat{a}_{\rm in}^{\prime}(t)\hat{b}_{\rm in}^{\prime\dagger}(t^{\prime})\rangle=m_{ab}\delta(t-t^{\prime})$ and $\langle\hat{a}_{\rm in}^{\prime}(t)\hat{b}_{\rm in}^{\prime}(t^{\prime})\rangle=-m_{ab}\delta(t-t^{\prime})$ where $m_{ab}=(G_{a}G_{b}\gamma/\omega_{m}^{2}\sqrt{\kappa_{a}\kappa_{b}})(2\bar{n}_{c}+1)$ \cite{Supple}. The output fields are similarly modified, with the indices ``in'' replaced by ``out'' and  $\hat{c}_{\rm out}=-\hat{c}_{\rm in}$ in this far off-resonant case. Note that the weak coupling assumption $|G_{a,b}|/\omega_{m}\ll1$, which allows the adiabatic elimination of the mechanical mode, also implies small values for the squeezing parameters $m_{a}$, $m_{b}$, and $m_{ab}$.

The polariton operators $\hat{A}$, $\hat{B}$ and their corresponding noise operators $\hat{A}_{\rm in}$, $\hat{B}_{\rm in}$ are readily obtained via a Bogoliubov transformation of the bare modes in the absence of dissipation. Assuming for simplicity $\kappa_{a}=\kappa_{b}=\kappa$, one then readily finds~\cite{Wang2012b}
\begin{equation}
\partial_{t}\hat{A} = (i\omega_{A}-\kappa)\hat{A}+\sqrt{2\kappa}\hat{A}_{\rm in},\label{eq:dA}
\end{equation}
and similarly for mode $B$, with $A\rightarrow B$.

Determining the conversion between the microwave signal and the optical field requires in general to solve the full Heisenberg-Langevin equations with time-dependent coefficients. But if one assumes perfect adiabaticity, one can use instead a much simplified effective two-sided cavity model. To single out the effect of the varying frequencies $\omega_{A,B}(t)$, we focus on the slowly varying envelopes $\widetilde{A}=\hat A e^{-i\omega_{A}t}$ and $\widetilde{B}=\hat B e^{-i\omega_{B}t}$. We also introduce a new operator for the symmetric superposition of the cavity modes, $\hat{V}=(\widetilde{A}+\widetilde{B})/\sqrt{2}$. From Eq.~(\ref{eq:dA}) we then have
\begin{equation}
\partial_{t}\hat{V}=-\kappa\hat{V}+\sqrt{\kappa}\widetilde{A}_{\rm in}+\sqrt{\kappa}\widetilde{B}_{\rm in},\label{eq:dV}
\end{equation}
reminiscent of the situation of a two-sided cavity~\cite{Walls1994} but with input field operators depending on $\Delta_{a}$. Specifically in the first stage of the detection sequence, $t<t_{0}$, we have $\widetilde{A}_{\rm in}\approx\hat{a}_{\rm in}^{\prime}e^{i\Delta_{a}^{\prime}t}$ and $\widetilde{B}_{\rm in}\approx\hat{b}_{\rm in}^{\prime}e^{i\Delta_{b}^{\prime}t}$, while in the third step, $t>t_{0}+\tau$, $\widetilde{A}_{\rm in}$ and $\widetilde{B}_{\rm in}$ are simply exchanged. In the intermediate second step, the adiabatic, essentially dissipation-free, evolution results in small phase shifts for the envelope operators, proportional to $\partial_{t}\omega_{A}$ and $\partial_{t}\omega_{B}$ for $\widetilde{A}$ and $\widetilde{B}$, respectively. In the case of perfect adiabaticity, we may neglect these shifts and thus obtain $\hat{V}(t_{0})=\hat{V}(t_{0} + \tau)$ \cite{Supple}. 

Summarizing, the full evolution of $\hat{V}$ for the three-step detection sequence is approximately described by the equation
\begin{equation}
\partial_{t}\hat{V}=-\kappa\hat{V}+\sqrt{\kappa}\hat{a}_{\rm in}^{\prime}e^{i\Delta_{a}^{\prime}t}+\sqrt{\kappa}\hat{b}_{\rm in}^{\prime}e^{i\Delta_{b}^{\prime}t}.
\end{equation}
With the boundary conditions of the two-sided cavity, $\hat{a}_{\rm out}^{\prime}e^{i\Delta_{a}^{\prime}t}+\hat{a}_{\rm in}^{\prime}e^{i\Delta_{a}^{\prime}t}=\sqrt{\kappa}\hat{V}$ and $\hat{b}_{\rm out}^{\prime}e^{i\Delta_{b}^{\prime}t}+\hat{b}_{\rm in}^{\prime}e^{i\Delta_{b}^{\prime}t}=\sqrt{\kappa}\hat{V}$~\cite{Walls1994}, this equation can be solved in the frequency domain to give
\begin{equation}
\hat{a}_{\rm out}^{\prime}(\omega-\Delta_{a}^{\prime})=\frac{\kappa\hat{b}_{\rm in}^{\prime}(\omega-\Delta_{b}^{\prime})-i\omega\hat{a}_{\rm in}^{\prime}(\omega-\Delta_{a}^{\prime})}{\kappa+i\omega}.
\end{equation}
Perfect conversion, $\hat{a}_{\rm out}^{\prime}(-\Delta_{a}^{\prime})=\hat{b}_{\rm in}^{\prime}(-\Delta_{b}^{\prime})$, occurs for $\omega=0$. Remembering that the optical and the microwave operators are expressed in rotating frames with respect to the pumping frequencies $\omega_{ap}$ and $\omega_{bp}$, this corresponds to the case where the frequency of the input microwave fields is $\omega_{s}=\omega_{b}-x_{c}g_{b0}-2G_{b}^{2}/\omega_{m}$ and the frequency of the output optical field is $\omega_{o} = \omega_{a}-x_c g_{a0}-2G_{a}^{2}/\omega_{m}$.

We introduce the mean photon numbers  of the optical and microwave modes 
\begin{eqnarray}
\bar{n}_o&=&\int d\omega |g(\omega)|^2\langle \hat{a}_{\rm out}^\dagger (\omega-\Delta_a^\prime)\hat{a}_{\rm out}(\omega-\Delta_a^\prime)\rangle\nonumber \\
\bar{n}_s&=&\int d\omega |g(\omega)|^2\langle \hat{b}_{\rm in}^\dagger (\omega-\Delta_b^\prime)\hat{b}_{\rm in}(\omega-\Delta_b^\prime)\rangle , 
\end{eqnarray}
where the mode filter functions $g(\omega)$ are sharply peaked around $\omega=0$. By assuming  detection and reception time windows $(\tau_d,\tau_r)\gg1/\kappa$ \cite{Vitali2004, Genes2008}, we find
\begin{equation}
\bar{n}_{o}=\bar{n}_{s}+\frac{(G_{b}^{2}+G_{a}^{2})\gamma}{\omega_{m}^{2}\kappa}(2\bar{n}_{c}+1),
\label{eq:nout}
\end{equation}
where we have taken into account the modified noise correlation of the optical and microwave cavities, and the effects of the mechanical noise are merged into the second term on the right-hand side. This is the central result of this Letter.

\emph{Sensitivity}.---Ignoring technical noise and assuming that the final optical detector is well characterized and has near unit quantum efficiency, we concentrate on the intrinsic sensitivity of the three-step conversion sequence. It is characterized primarily by the microwave-to-optical conversion efficiency, the effects of quantum and thermal noise, and the dead time required to reset the resonators between measurements. Perfect adiabatic conversion requires interaction times $\kappa \ll 1/\tau \ll 4|G_{a}G_{b}|/\omega_{m} \ll \omega_m$, and the dead time to reset the resonators is of the order of $1/\kappa$.  Quantum and thermal noise result in a dark-count rate that also impacts the figure of merit of the detector; see Eq.~(\ref{eq:nout}). A high-$Q$ and ultracold mechanical oscillator can significantly suppress these sources of noise. 

As an example we consider an optomechanical resonator with high mechanical frequency $\omega_m=2\pi\times4$GHz and quality factor $Q=87\times10^3$, which results in $\gamma=2\pi\times46$ kHz and $\bar{n}_c= 72$ for a temperature $T=14$K \cite{Hill2012}.  Because of the large detunings considered here, we find, however, that the mechanical noise only adds a contribution of $0.06$  to $\bar n_o$. The level of thermal microwave noise that feeds into $\bar n_s$ can be managed by cooling the microwave cavity to cryogenic temperatures. For a microwave cavity frequency $\omega_{b}=2\pi\times300$ GHz and temperature $T_b=300$ K, we have $\bar{n}_{s}=20$, but for $T_b=3$ K, $\bar{n}_{s}$ is reduced to $0.008$. Finally, we assume linear optomechanical coupling strengths $G_a=-2\pi\times200$ MHz and $G_b=2\pi\times300$ MHz, respectively, giving an effective interaction strength $2G_aG_b/\omega_m=-2\pi\times30$ MHz. We also set the same decay rate for both cavities, $\kappa=2\pi\times850$ kHz.  These parameters fulfill the condition for adiabaticity of the conversion and result in a dead time of the order of $100$~ns.  These estimates indicate that the detector should be able to operate reliably at or below the single-photon level.

\emph{Conclusion}.---We have proposed and analyzed a time-gated microwave detection scheme based on the control of polaritons in a hybrid optomechanical system. In contrast to resonant schemes that focus on high fidelity quantum state transfer \cite{Barzanjeh2012, Wang2012, Tian2012, Zhangqi2014}, the dual optomechanical cavity detector is driven by a heterodynelike pumping and operates on the far-off sideband resonant regime to minimize pump and  mechanical noise, thereby offering the potential to reliably detect very feeble microwave fields.  Importantly, that nonresonant approach does not preserve the quantum state of the microwave field. Rather, it detects the signal entering the microwave resonator in a time determined by its decay time $1/\kappa_b$ just before transfer to the optical domain.

We thank S. Singh, H. Seok, P. Treutlein, H. Metcalf, R. Dehghannasiri and A. A. Eftekhar for helpful discussions. This work was supported by the National Basic Research Program of China Grant No. 2011CB921604, the NSFC Grants No.~11204084, No. 91436211, No. 11234003, and No. 11304072, the SRFDP Grant No.~20120076120003, the SCST  Grant No.~12ZR1443400, the DARPA QuASAR and ORCHID programs through grants from AFOSR and ARO, the U.S. Army Research Office, and  NSF. Y. D. is supported in part by the Hongzhou-city Quantum Information and Quantum Optics Innovation Research Team.


\begin{widetext}
\clearpage
\begin{center}
\textbf{\large Supplemental Materials: Proposal for an Optomechanical Microwave Sensor at the Subphoton level}\\
\bigskip
Keye Zhang$^{1,2}$, Francesco Bariani$^2$, Ying Dong$^{2,3}$, Weiping Zhang$^1$, Pierre Meystre$^{2}$\\
\bigskip
       \footnotesize {\textit{$^1$Quantum Institute for Light and Atoms, State Key Laboratory of Precision Spectroscopy, Department of Physics, East
China Normal University, Shanghai, 200241, China }\\
       \textit{$^2$B2 Institute, Department of Physics and College of Optical Sciences,
University of Arizona, Tucson, Arizona 85721, USA}\\
\textit{$^3$Department of Physics, Hangzhou Normal University, Hangzhou, Zhejiang 310036, China}}

\end{center}
\end{widetext}
\setcounter{equation}{0}
\setcounter{figure}{0}
\setcounter{table}{0}
\setcounter{page}{1}
\makeatletter
\renewcommand{\theequation}{S\arabic{equation}}
\renewcommand{\thefigure}{S\arabic{figure}}
\renewcommand{\bibnumfmt}[1]{[S#1]}
\renewcommand{\citenumfont}[1]{S#1}

\section{Intermode scattering}

This section discusses intermode scattering in the optical and microwave resonators  and justifies its neglect under the conditions considered in the main text.

We consider a system comprised of a two-mode optical cavity with bosonic annihilation operators $\hat a$ and $\hat a_{p}$ and a two-mode microwave cavity with annihilation operators $\hat b$ and $\hat b_{p}$, optomechanically coupled by a common radiation pressure driven end-mirror. The center-of-mass motion of that mirror is quantized and characterized by the annihilation operator $\hat c$. The interaction between the three subsystems is described by the optomechanical Hamiltonian
\begin{eqnarray}
V &=& \hbar g_{a0}(\hat a+\hat a_p)^\dagger(\hat a+\hat a_p)(\hat c + \hat c^\dagger) \nonumber\\
&+& \hbar g_{b0}(\hat b+\hat b_p)^\dagger(\hat b+\hat{b}_p)(\hat{c}+\hat{c}^{\dagger}),
\end{eqnarray}
where $g_{a0}$ and $g_{b0}$ are the optical and microwave single-photon optomechanical coupling strengths, respectively. We assume that $g_{a0}$ and $g_{b0}$ are real but with opposite signs. The cavity modes $\hat{a}_{p}$ and $\hat{b}_{p}$ are resonantly driven by two classical fields at frequencies $\omega_{a,p}$ and $\omega_{b,p}$, respectively, so that in the frame rotating at the pumping frequencies, the Heisenberg-Langevin equations of motion have the form
\begin{widetext}
\begin{eqnarray}
\partial_{t}\hat{a} & = & (i\Delta_a-\kappa_a)\hat{a}-ig_{a0}(\hat a+\hat a_p)(\hat{c}+\hat{c}^{\dagger})+\sqrt{2\kappa_{a}}\hat{a}_{\rm in},\label{eq:Sda}\\
\partial_{t}\hat{b} & = & (i\Delta_{b}-\kappa_{b})\hat{b}-ig_{b0}(\hat{b}+\hat{b}_{p})(\hat{c}+\hat{c}^{\dagger})+\sqrt{2\kappa_{b}}\hat{b}_{\rm in},\label{eq:Sdb}\\
\partial_{t}\hat{c} & = & (-i\omega_{m}-\gamma)\hat{c}-ig_{a0}(\hat{a}+\hat{a}_{p})^{\dagger}(\hat{a}+\hat{a}_{p})-ig_{b0}(\hat{b}+\hat{b}_{p})^{\dagger}(\hat{b}+\hat{b}_{p})+\sqrt{2\gamma}\hat{c}_{\rm in},\label{eq:Sdc}\\
\partial_{t}\hat{a}_{p} & = & -\kappa_{a}\hat{a}_{p}-ig_{a0}(\hat{a}+\hat{a}_{p})(\hat{c}+\hat{c}^{\dagger})+\eta_{a}+\sqrt{2\kappa_{a}}\hat{a}_{p, \rm in},\label{eq:Sdap}\\
\partial_{t}\hat{b}_{p} & = & -\kappa_{b}\hat{b}_{p}-ig_{b0}(\hat{b}+\hat{b}_{p})(\hat{c}+\hat{c}^{\dagger})+\eta_{b}+\sqrt{2\kappa_{b}}\hat{b}_{p, \rm in}.\label{eq:Sdbp}
\end{eqnarray}
\end{widetext}
Here  $\Delta_a=\omega_{a,p}-\omega_{a}$ and $\Delta_{b}=\omega_{b,p}-\omega_{b}$ are the optical and the microwave cavity-pumping detunings, $\eta_a$ and $\eta_b$ represent the strength of the pump fields, and the dissipation effects are described by the decay rates $\kappa_a$, $\kappa_b$ and $\gamma$,  with corresponding noise
operators labeled by the subscript ``in''. 

For strong pumping one can linearize the dynamics around the classical steady state with the substitutions $\hat{a}\rightarrow\alpha+\hat{a}$, $\hat{a}_{p}\rightarrow\alpha_{p}+\hat{a}_{p}$, $\hat{b}\rightarrow\beta+\hat{b}$, $\hat{b}_{p}\rightarrow\beta_{p}+\hat{b}_{p}$, $\hat{c}\rightarrow\mathcal{C}+\hat{c}$, with steady values given to leading order by
\begin{eqnarray}
\beta_p & \approx & \frac{\eta_b}{\kappa_b+ig_{b0}x_c},\\
\beta  &\approx & \frac{g_{b0}\beta_p x_c}{\Delta_b},\\
\alpha_p & \approx & \frac{\eta_a}{\kappa_a+ig_{a0}x_c},\\
\alpha  &\approx & \frac{g_{a0}\alpha_p x_c}{\Delta_a},\\
x_c =\mathcal{C}+\mathcal{C}^*  &\approx& -\frac{g_{b0}|\beta_p|^2+g_{a0}|\alpha_p|^2}{\omega_m},\label{eq:Sxc}
\end{eqnarray}
and we have assumed 
\begin{equation}
\omega_m,|\Delta_a|,|\Delta_b|\gg\kappa_a,\kappa_b,\gamma.
\label{rates}
\end{equation}

 Due to the large detuning from the pump, the steady amplitudes $\alpha$ and $\beta$ of the cavity modes $\hat{a}$ and $\hat{b}$ can be ignored when compared to the amplitudes $\alpha_p$ and $\beta_p$ of the driven cavity modes $\hat{a}_p$ and $\hat{b}_p$. To lowest order in the quantum fluctuations about the classical steady state, Eqs. (\ref{eq:Sda})-(\ref{eq:Sdbp}) reduce then to 
\begin{widetext}
\begin{eqnarray}
\partial_{t}\hat a & = & (i\Delta_a-ig_{a0}x_c-\kappa_a)\hat a-iG_a(\hat c+\hat c^\dagger)-ig_{a0}x_c\hat a_p -ig_{a0}(\hat c+\hat c^\dagger)\hat a_p +\sqrt{2\kappa_a}\hat a_{\rm in},\label{eq:Sda1}\\
\partial_{t}\hat b & = & (i\Delta_b-ig_{b0}x_c-\kappa_b)\hat b-iG_b(\hat c+\hat c^{\dagger})-ig_{b0}x_c\hat b_p -ig_{b0}(\hat c+\hat c^\dagger)\hat b_p +\sqrt{2\kappa_b}\hat b_{\rm in},\label{eq:Sdb1}\\
\partial_{t}\hat c & = & (-i\omega_m-\gamma)\hat c-i[G_a(\hat a+\hat a_p) + h.c.]-i[G_b(\hat b+\hat b_p) + h.c.]+\sqrt{2\gamma}\hat c_{\rm in},\label{eq:Sdc1}\\
\partial_{t}\hat a_p & = & (-ig_{a0}x_c-\kappa_a)\hat a_p-iG_a(\hat c+\hat c^\dagger)-ig_{a0}x_c\hat a-ig_{a0}(\hat c+\hat c^\dagger)\hat a+\sqrt{2\kappa_a}\hat a_{p, \rm in},\label{eq:Sdap1}\\
\partial_{t}\hat b_p & = & (-ig_{b0}x_c-\kappa_b)\hat b_p-iG_b(\hat c+\hat c^\dagger)-ig_{b0}x_c\hat b -ig_{b0}(\hat c+\hat c^\dagger)\hat b+\sqrt{2\kappa_b}\hat b_{p, \rm in}.\label{eq:Sdbp1}
\end{eqnarray}
\end{widetext}
The second terms on the right-hand side of Eqs.~(\ref{eq:Sda1}) and (\ref{eq:Sdb1}) describe the usual linearly enhanced  optomechanical coupling, with $G_{b}=\beta_{p}g_{b0}$ and $G_{a}=\alpha_{p}g_{a0}$, taken to be real in the following. They correspond to the first term of the Hamiltonian $V_{3m,\rm eff}$ in the main text. 

The third and fourth terms, proportional to $\hat a_p$ and $\hat b_p$,  account for the scattering  of the classically driven cavity modes into the modes $\hat a$ and $\hat b$. 
First, we note that it is possible to adjust the pumping fields in such a way that the mean radiation pressure forces from the optical and microwave fields cancel out, i.e., $x_c\sim 0$, see Eq.~(\ref{eq:Sxc}) \footnote{Note that $x_c$ is a dimensionless quantity}. It follows that the scattering of the quantum fluctuations of the classically driven modes can safely be neglected. 

The contribution of the second scattering term, proportional to $(\hat c + \hat c^\dagger)$, is estimated by 
\begin{equation}
S \propto \int_{0}^t dt' e^{(i\Delta_b-\kappa_b)(t-t^\prime)}[\hat c(t^\prime)+\hat c^\dagger(t^\prime)]\hat b_p(t^\prime).
\end{equation}
From Eqs.~(\ref{rates}) and (\ref{eq:Sdbp1}), and in the usual situation where the single-photon optomechanical couplings are very weak, and noting that $\hat c$ is fast-oscillating at the frequency $\omega_m$, we have that  $\hat b_p$ is slowly-varying and can be safely moved out of the integral.  Isolating the fast varying contribution to $\hat c$ with   $\hat c=\tilde c e^{-i\omega_m t}$ and moving its slowly-varying envelope $\tilde c$ outside of the integral gives then
\begin{widetext}
\begin{equation}
S=-ig_{b0}\hat b_p\left[\left(\frac{e^{-i\omega_m t}-e^{(i\Delta_b-\kappa_a)t}}{-i\omega_m-i\Delta_b+\kappa_b}\right)\tilde c 
+ \left(\frac{e^{i\omega_m t}-e^{(i\Delta_b-\kappa_a)t}}{i\omega_m-i\Delta_b+\kappa_b}\right)\tilde c^\dagger\right],
\end{equation}
\end{widetext}
This shows that as expected intermode scattering is maximized in the resonant case $\Delta_b=\pm\omega_m$. But in the off-resonant case $|\Delta_b| < \omega_m$ , neglecting the fast-oscillating term $e^{-i\omega_m t}$ we have
\begin{equation}
S \approx \frac{g_{b0}}{\omega_m}\hat b_p(\tilde c^\dagger-\tilde c)e^{(i\Delta_b-\kappa_b)t}
\end{equation}
and the contribution of the scattering from the ancillary mode to the average population of the mode $\hat b$ is approximatively
\begin{equation}
\langle\hat b^\dagger\hat b\rangle\thicksim\frac{g_{b0}^2}{\omega_m^2}\bar n_{bp}(2\bar n_c+1).
\end{equation}
In this last step we have  neglected possible correlations between the pumped mode and the phonon mode and introduced the thermal mean populations $\bar n_{bp}$ of the ancillary cavity mode and $\bar n_c$ of the mechanics. We compare these expressions to the noise deriving from the radiation pressure off-resonant coupling, see Eq.(6) in the main text and the discussion of the next section, 
\begin{equation}
\langle\hat b^\dagger\hat b\rangle\thicksim\frac{g_{b0}^2|\beta_p|^2}{\omega_m^2}\frac{\gamma}{\kappa}(2\bar n_c+1).
\end{equation}
It is apparent that for $|\beta_p|^2\gg\bar n_{bp}$, the situation considered here,  scattering noise represent a small correction only. Similar results can also be derived for the optical side of the double resonator system.

Finally, comparing Eqs.~(\ref{eq:Sda1})-(\ref{eq:Sdbp1}) we have that for $|g_{a0}x_{c}|,|g_{b0}x_{c}|\ll|\Delta_{a}|,|\Delta_{b}|,\omega_{m}$ the characteristic frequencies of the classically driven modes $\hat a_{p}$ and $\hat b_{p}$ are far from those of the modes $\hat a$, $\hat b$, and $\hat c$ . In that limit these modes are effectively decoupled from the other three modes.  These arguments justify ignoring the second and the last term of the full Hamiltonian $V_{3m,\rm eff}$ in the main text. Neglecting the fluctuation of the pumped modes results in the linearized effective three-mode Hamiltonian
\begin{eqnarray}
H&=&\hbar\omega_m\hat{c}^{\dagger}\hat{c}-\hbar\Delta_{a}\hat{a}^{\dagger}\hat{a}-\hbar\Delta_{b}\hat{b}^{\dagger}\hat{b}\nonumber \\
&+&\hbar G_{a}(\hat{a}+\hat{a}^{\dagger})(\hat{c}+\hat{c}^{\dagger})\nonumber \\
&+& \hbar G_{b}(\hat{b}+\hat{b}^{\dagger})(\hat{c}+\hat{c}^{\dagger})+H_{\kappa},\label{eq:SH3}
\end{eqnarray}
where 
$$
\Delta_a=\omega_{a,p}-\omega_a+x_xg_{a0},
$$
$$\Delta_b=\omega_{b,p}-\omega_b+x_cg_{b0},
$$
 and the Hamiltonian $H_{\kappa}$ accounts for the effects of dissipation.

\section{Adiabatic elimination of the phonon mode}

This section presents details of the adiabatic elimination of the phonon mode. 

The effective three-mode Hamiltonian (\ref{eq:SH3}) yields the Heisenberg-Langevin equations
\begin{widetext}
\begin{eqnarray}
\partial_{t}\hat{a} & = & (i\Delta_{a}-\kappa_{a})\hat{a}-iG_{a}(\hat{c}+\hat{c}^{\dagger})+\sqrt{2\kappa_{a}}\hat{a}_{\rm in},\label{eq:Sda2}\\
\partial_{t}\hat{b} & = & (i\Delta_{b}-\kappa_{b})\hat{b}-iG_{b}(\hat{c}+\hat{c}^{\dagger})+\sqrt{2\kappa_{b}}\hat{b}_{\rm in},\\
\partial_{t}\hat{c} & = & (-i\omega_{m}-\gamma)\hat{c}-iG_{a}(\hat{a}+\hat{a}^{\dagger})-iG_{b}(\hat{b}+\hat{b}^{\dagger})+\sqrt{2\gamma}\hat{c}_{\rm in},\label{eq:Sdc2}
\end{eqnarray}
\end{widetext}
where the input fields, $\hat{a}_{\rm in}$, $\hat{b}_{\rm in}$, and $\hat{c}_{\rm in}$ are assumed to be well approximate as white noise sources with zero mean,
\begin{eqnarray}
\langle\hat{a}_{\rm in}^{\dagger}(t)\hat{a}_{\rm in}(t^{\prime})\rangle & = & \bar{n}_{a}\delta(t-t^{\prime}),\\
\langle\hat{b}_{\rm in}^{\dagger}(t)\hat{b}_{\rm in}(t^{\prime})\rangle & = & \bar{n}_{b}\delta(t-t^{\prime}),\\
\langle\hat{c}_{\rm in}^{\dagger}(t)\hat{c}_{\rm in}(t^{\prime})\rangle & = & \bar{n}_{c}\delta(t-t^{\prime}).\label{cindcin}
\end{eqnarray}
If the phonon frequency $\omega_{m}$ is significantly larger than any other frequency scale we can adiabatically eliminate the dynamics of the mechanical mode, thereby reducing the description of the system to a two-mode model. Specifically, from Eq.~(\ref{eq:Sdc2}) we have
\begin{widetext}
\begin{eqnarray}
\hat c (t)&=&\hat c(0)e^{(-i\omega_m-\gamma)t}-i\int_0^te^{(-i\omega_m-\gamma)(t-t^\prime)}[G_a\hat X_a(t^\prime)+G_b\hat X_b(t^\prime)]dt' \nonumber \\
&+ &\sqrt{2\gamma}\int_0^t e^{(-i\omega_m-\gamma)(t-t^\prime)}\hat c_{\rm in}(t^\prime)dt^\prime,
\end{eqnarray}
where $\hat X_a\equiv \hat a+\hat a^\dagger$ and $\hat X_b\equiv \hat b+\hat b^\dagger $. Substituting that expression into Eq.~(\ref{eq:Sda2}) gives
\begin{eqnarray}
\partial_{t}\hat{a} & = & (i\Delta_a-\kappa_a)\hat a -\int_{0}^{t}e^{(-i\omega_m-\gamma)(t-t^\prime)}[G_a^2\hat X_a(t^\prime)+G_a G_b\hat X_b(t^\prime)]dt^\prime \label{eq:Sda3}\\
 &+ &\int_{0}^{t}e^{(i\omega_m-\gamma)(t-t^\prime)}[G_a^{2}\hat X_a(t^\prime)+G_a G_b\hat X_b(t^\prime)]dt' 
 -iG_a [\hat c(0)e^{(-i\omega_m -\gamma)t}+h.c.]+\sqrt{2\kappa_a}\hat a_{\rm in}^\prime,\nonumber 
\end{eqnarray}

where the noise
\begin{equation}
\hat{a}_{\rm in}^{\prime}=\hat{a}_{\rm in} -iG_{a}\sqrt{\frac{\gamma}{\kappa_{a}}}\left[\int_{0}^{t}e^{(-i\omega_{m}-\gamma)(t-t^{\prime})}\hat{c}_{\rm in}(t^{\prime})dt^{\prime}+h.c.\right]
\label{eq:Snoise}
\end{equation}
\end{widetext}
is characterized in general by a  colored spectrum \cite{Chiocchetta14S, Singh12S}. 

We assume that the quadratures $\hat X_{a,b}$ are slowly varying and move them out of the integral. The second term of Eq. (\ref{eq:Sda3}) becomes
\begin{equation}
\frac{1}{\omega_m}\left(- i + \frac{\gamma}{\omega_m}\right)(G_a^2\hat X_a+G_a G_b\hat X_b),\label{eq:S2nd}
\end{equation}
where we have neglected fast-oscillating terms proportional to $e^{-i\omega_{m}t}$ and expanded the result to the first order with respect to the small quantity $\gamma/\omega_{m}$. Similarly, the third term may be approximated as
\begin{equation}
\frac{1}{\omega_m}\left(- {i} - \frac{\gamma}{\omega_m}\right)(G_a^2\hat X_a+G_a G_b\hat X_b).\label{eq:S3rd}
\end{equation}
The modifications of the linewidth of the optical mode $\kappa_{a}$ due to Eqs. (\ref{eq:S2nd}) and (\ref{eq:S3rd}) cancel each other out. 

The fourth term, involving $\hat c(0)$ and $\hat c^\dagger(0)$, is a memory effect that describes the dependence of the cavity field operator at time $t$ on the initial phonon operators,
\begin{equation}
  \hat a(t)\sim \frac{G_a[e^{(-i\omega_m-\gamma)t}-e^{(-i\Delta_a-\kappa_a)t}]}{\omega_m-\Delta_a-i\gamma+i\kappa_a }\hat c (0)-h.c.
\end{equation}
so that the mean optical intensity $\langle\hat a^\dagger(t)\hat a(t)\rangle$  contains a contribution from the initial inter-mode scattering. Since it includes a term that anti-normally ordered correlation function $\langle\hat c(0)\hat c^\dagger(0)\rangle$ that contribution persists even if the fast-oscillating contributions proportional to $e^{\pm i\omega_m t}$ are ignored and the mechanical oscillator is initially in its ground state. However for long enough times it does become negligible due to the exponential decay factors $e^{-2\kappa_a t}$ and $e^{-2\gamma t}$. It can therefore be ignored provided that the ``receiving time'' window is much longer than {\em both} the cavity decay time $1/\kappa_a$ and the mechanical damping time $1/\gamma$. 

We now turn to the contribution of the phonon noise term in Eq.~(\ref{eq:Snoise}). The precise form of its contribution to the normal order noise correlation $\langle\hat a_{\rm in}^{\prime\dagger}(t_1)\hat a_{\rm in}^{\prime}(t_2)\rangle$ is given by a double integral
\begin{widetext}
\begin{equation}
m(t_1, t_2)=\frac{G_{a}^{2}\gamma}{\kappa_{a}}\int_{0}^{t_{1}}dt\int_{0}^{t_{2}}dt^{\prime}e^{i\omega_{m}(t_{1}-t_{2}+t^{\prime}-t)}e^{-\gamma(t_{1}+t_{2}-t-t^{\prime})}\langle\hat{c}_{\text{in}}^{\dagger}(t)\hat{c}_{\text{in}}(t^{\prime})\rangle+h.c.
\end{equation}
With the correlation function~(\ref{cindcin}) and for $t_2>t_1$ this gives
\begin{equation}
m(t_1,t_2) \approx m(\tau)=\frac{G_{a}^{2}\bar{n}_{c}}{2\kappa_{a}}e^{(-i\omega_{m}-\gamma)\tau}+\frac{G_{a}^{2}(\bar{n}_{c}+1)}{2\kappa_{a}}e^{(i\omega_{m}-\gamma)\tau},
\label{Ntau}
\end{equation}
\end{widetext}
where $\tau\equiv t_2-t_1$ and we have discarded  terms proportional to $e^{-\gamma(t_{1}+t_{2})}$ for times to $\{t_1,t_2\}\gg 1/\gamma$ , so that $m$ is a function of $\tau$ only. This approximate form is consistent with the neglect of memory effects just discussed. For the case $t_2<t_1$ we will obtain a similar expression with $\gamma$ replaced by $-\gamma$ \cite{Singh12S}. 

The correlation $m(\tau)$ vanishes rapidly over the characteristic time scale of the cavity mode dynamics ($1/\Delta_a$) due to the the fast oscillating factors $e^{\pm i\omega_{m}\tau}$ for $\omega_m\gg\Delta_a$, except for $\tau=0$ \cite{LouisellS}. The phonon noise term in Eq.~(\ref{eq:Snoise}) can then be approximated by a $\delta$-correlated noise operator with $m(\tau)\approx m_a\delta(\tau)$ as far as the cavity mode dynamics is concerned, with $m_a$ given by the integral of $m(\tau)$ over the full temporal domain except for the pole at $\tau=0$,
\begin{eqnarray}
m_a & = & \int_{-\infty}^{0^-}m(\tau)d\tau+\int_{0^+}^{+\infty}m(\tau)d\tau\nonumber \\
 & \approx & \frac{G_{a}^{2}\gamma}{\kappa_a \omega_m^2}(2\bar{n}_{c}+1).
\end{eqnarray}
We can invoke the same approximation for other  noise correlations, as for example $\langle\hat a_{\rm in}^\prime(t)\hat a_{\rm in}^\prime(t^\prime)\rangle$. This is equivalent  to performing an adiabatic approximation on Eq.~(\ref{eq:Snoise}), which results in 
\begin{equation}
\hat a_{\rm in}^\prime \approx \hat a_{\rm in}+\frac{G_a}{\omega_m}\sqrt{\frac{\gamma}{\kappa_a}}(\hat c_{\rm in}-\hat c_{\rm in}^\dagger)
\end{equation}
with the squeezed reservoir correlation functions
\begin{eqnarray}
\langle\hat a_{\rm in}^{\prime\dagger}(t)\hat a_{\rm in}^\prime(t^\prime)\rangle & = & (\bar n_a+m_a)\delta(t-t^\prime), \label{adain}\\
\langle\hat a_{\rm in}^\prime(t)\hat a_{\rm in}^\prime(t^\prime)\rangle & = & -m_a\delta(t-t^\prime).
\label{squeeze}
\end{eqnarray}
We emphasize that these approximations rely on the frequency difference between the cavity mode and the phonon mode ($\omega_m\gg\Delta_a$) being large, and also that the main contribution of the phonon noise correlation to the cavity mode is at frequencies close to $\Delta_a$, as is confirmed by the noise analysis of the next section.

Following similar steps for the microwave mode $\hat{b}$  we finally obtain the two-mode Heisenberg-Langevin equations
\begin{widetext}
\begin{eqnarray}
\partial_t\hat a & = & (i\Delta_a-\kappa_a)\hat a+i\frac{2G_a^2}{\omega_m}(\hat a+\hat a^\dagger)+i\frac{2G_a G_b}{\omega_m}(\hat b+\hat b^\dagger)+\sqrt{2\kappa_a}\hat a_{\rm in}^\prime,\label{eq:Sda4}\\
\partial_t\hat b & = & (i\Delta_b-\kappa_b)\hat b+i\frac{2G_b^2}{\omega_m}(\hat b+\hat b^\dagger)+i\frac{2G_a G_b}{\omega_m}(\hat a+\hat a^\dagger)+\sqrt{2\kappa_b}\hat b^\prime_{\rm in}.\label{eq:Sdb4}
\end{eqnarray}
\end{widetext}
with the reservoir correlation functions
\begin{eqnarray}
\langle\hat b_{\rm in}^{\prime\dagger}(t)\hat b_{\rm in}^\prime(t^\prime)\rangle & = & (\bar n_b+m_b)\delta(t-t^\prime), \label{bdbin}\\
\langle\hat b_{\rm in}^\prime(t)\hat b_{\rm in}^\prime(t^\prime)\rangle & = & -m_b\delta(t-t^\prime),
\end{eqnarray}
and the cross correlation functions
\begin{eqnarray}
\langle\hat a_{\rm in}^{\prime\dagger}(t)\hat b_{\rm in}^\prime(t^\prime)\rangle & = & m_{ab}\delta(t-t^\prime),\\
\langle\hat a_{\rm in}^\prime(t)\hat b_{\rm in}^\prime(t^\prime)\rangle & = & -m_{ab}\delta(t-t^\prime). \label{ab}
\end{eqnarray}
Here $m_b=(G_b^2\gamma/\kappa_a \omega_m^2)(2\bar n_c+1)$ and $m_{ab}=(G_aG_b\gamma/\sqrt{\kappa_a\kappa_b}\omega_m^2)(2\bar n_c+1)$.

\section{Noise analysis in the frequency domain}

This section presents a frequency domain noise analysis of the detector during the ``receiving'' time window. This permits to perform a comparison between the full five-mode model  and the effective two-mode model aimed at confirming the validity of the adiabatic approximation used in the main text.

From Section I of the Supplemental Material the five-mode coupled equations are linear and can therefore be solved in frequency domain. We proceed by expressing Eqs.~(\ref{eq:Sda1})-(\ref{eq:Sdbp1}) in matrix form as
\begin{equation}
\dot \mathcal O(t)=\mathcal M \mathcal O(t)+\mathcal O_{\rm in}(t),
\label{matrixform}
\end{equation}
where 
\begin{equation}
\mathcal{O}(t)=\left(\begin{array}{c}
\hat{o}(t)\\
\hat{o}^{\dagger}(t)
\end{array}\right),\:\hat{o}(t)=\left(\begin{array}{c}
\hat{a}(t)\\
\hat{b}(t)\\
\hat{c}(t)\\
\hat{a}_{p}(t)\\
\hat{b}_{p}(t)
\end{array}\right),
\end{equation}
and where $\mathcal O_{\rm in}(t)$ accounts for the contribution of the noise operators, with correlation functions
\begin{equation}
\langle \hat o^\dagger_{\rm in}(t)\hat o_{\rm in}(t')\rangle=\bar n_o\delta(t-t') .
\label{ocorrt}
\end{equation}
Introducing the compact notation
\begin{equation}
\mathcal{O}(\omega)=\frac{1}{\sqrt{T}}\int_{-T/2}^{T/2}e^{i\omega t}\mathcal{O}(t)dt=\left(\begin{array}{c}
\hat{o}(\omega)\\
\hat{o}^{\dagger}(-\omega)
\end{array}\right),
\end{equation}
where $T$ the measurement time, we can make the substitution $\mathcal{O}(t)\rightarrow\mathcal{O}(\omega)$ and $\dot\mathcal{O}(t)\rightarrow -i\omega\mathcal{O}(\omega)$ in Eq(\ref{matrixform}), resulting in a matrix equation that can be solved by simple linear algebra. Its solution can be written in the form
\begin{equation}
\mathcal O(\omega)=\mathcal X(\omega)\mathcal O_{\rm in}(\omega),
\label{frequencysolution}
\end{equation}
where $\mathcal X(\omega)$ is the matrix of response functions and $\mathcal O_{\rm in}(\omega)$ includes all frequency domain noise operators, with correlation functions obtained from Eq.~(\ref{ocorrt}) as
\begin{equation}
\langle \hat o^\dagger_{\rm in}(\omega)\hat o_{\rm in}(\omega')\rangle=\bar n_o\delta(\omega+\omega') .
\label{frequencycorrelation}
\end{equation}
The power spectral density for mode $\hat o$ is~\cite{GardinerS}
\begin{equation}
\mathcal S_{oo}(\omega)=\int_{-\infty}^{\infty}\langle \hat o^\dagger(\omega)\hat o(-\omega')\rangle d\omega',
\end{equation}
and its mean excitation number is
\begin{equation}
n_o=\int_{-\infty}^{\infty}\mathcal S_{oo}(\omega) d\omega.
\end{equation}

\begin{figure*}
\includegraphics[width = \textwidth]{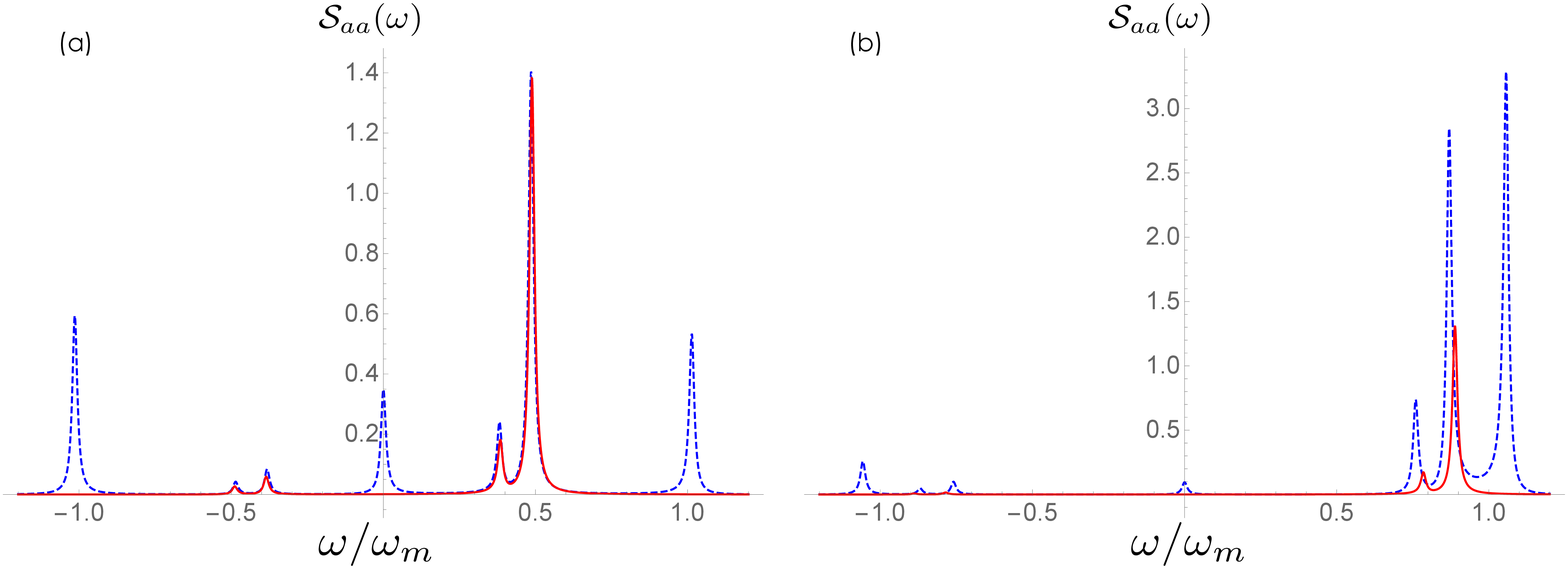}
\caption{(Color online) (a) Power spectral density of the optical mode $\hat a$ obtained from the original five-mode model (blue dashed curve) and from the effective two-mode model under the adiabatic approximation (red solid curve) with parameters $\Delta_a=-0.5$, $\Delta_b=-0.4$, $G_a=-G_b=0.08$, $\kappa_a=\kappa_b=\gamma=0.01$, $\bar n_a=\bar n_{ap}=0$, $\bar n_b=\bar n_{bp}=0.04$, and $\bar n_c=0.1$ ; (b) Same parameters as in (a), but with $\Delta_a=-0.9$, $\Delta_b=-0.8$. All quantities are normalized to the phonon frequency $\omega_m$.\label{fig:spectrum}}
\end{figure*}
Taking the optical mode $\hat a$ as an example, its spectal density $\mathcal S_{aa}(\omega)$, as well as the mean photon number $n_a$,  depend on the response function $\mathcal X(\omega)$ and the frequency noise correlations of all  five modes. Following  similar steps we can also obtain $\mathcal S_{aa}(\omega)$ for the effective two-mode model of Eqs.~(\ref{eq:Sda4}) and (\ref{eq:Sdb4}) with the modified noise correlation functions of Eqs. (\ref{adain}), (\ref{squeeze}), and (\ref{bdbin})-(\ref{ab}). 

Figure~\ref{fig:spectrum} compares the power spectral density $\mathcal S_{aa}(\omega)$ obtained from the 5-mode and the 2-mode models in two cases for a red-detuned laser impinging on the cavity ($\Delta_a<0$). In case (a), $\omega_m\gg|\Delta_{a,b}|$,  $\mathcal S_{aa}(\omega)$ is characterized by a main peak at the positive frequency $-\Delta_a$, a very weak peak at $\Delta_a$,  as well as additional  peaks at $\omega = \pm \Delta_b$, $\omega = \pm \omega_m$, and $\omega = 0$ representing contributions from the microwave mode, phonon mode, and the pumping mode, respectively. In contrast the approximate two-mode adiabatic result (red solid curve) has only the main peak and the peak from the microwave mode, but the key point is that they are in perfect agreement with the corresponding part of five-mode spectrum. That means that when considering signals at frequencies close to $-\Delta_{a,b}$ the effective adiabatic two-mode model is a good approximation. In contrast, case (b) where $\omega_m\sim|\Delta_{a,b}|$, illustrates a situation where the conditions required to perform the adiabatic approximation are not satisfied. In that case the power spectral density from the two-mode model is quite different from its five-mode counterpart, even for frequencies close to $-\Delta_a$.

\section{Effective two-sided cavity model}

This section provides more details about the analogy between the theoretical description of the proposed detector and the two-side cavity model of quantum optics.

From the main text, the Heisenberg-Langevin equations for the normal modes $\hat{A}$
and $\hat{B}$ read 
\begin{eqnarray}
\partial_t\hat A & = & (i\omega_A-\kappa)\hat A+\sqrt{2\kappa}\hat A_{\rm in},\\
\partial_t\hat B & = & (i\omega_B-\kappa)\hat B+\sqrt{2\kappa}\hat B_{\rm in}.
\end{eqnarray}
We can combine them into the equation for a symmetric superposition of the two slowly varying envelops
\begin{equation}
\hat V=\frac{\hat A e^{-i\omega_A t}+\hat B e^{-i\omega_B t}}{\sqrt{2}},
\end{equation}
with two input fields $\widetilde A_{\rm in}=\hat A_{\rm in}e^{-i\omega_A t}$
and $\widetilde B_{\rm in}=\hat B_{\rm in}e^{-i\omega_B t}$,
\begin{equation}
\partial_t\hat V=-\kappa\hat V+\sqrt{\kappa}\widetilde A_{\rm in}+\sqrt{\kappa}\widetilde B_{\rm in}.\label{eq:SdVdt}
\end{equation}
This equation is formally identical to the description of a two-sided cavity with symmetric decays rate $\kappa/2$ on both cavity sides~\cite{Walls94S}. It is easily solved in the frequency domain, but the simple solution must be adapted to the present situation since the form of the normal-mode input-field operators and their frequencies are now time dependent.

During the first stage of the detection sequence, $t<t_0$, the normal mode $A$ is optical-like. We have $\hat A_{\rm in}=\hat a_{\rm in}^\prime$ and $\omega_A=-\Delta_a^\prime$ with $\hat a_{\rm in}^\prime$ and $\Delta_a^\prime$ the modified optical input operator and detuning given in Eq. (6) in the main text. Furthermore, 
\begin{equation}
\widetilde A_{\rm in}=\tilde a_{\rm in}^\prime e^{-i\omega_o t},\label{eq:SAint}
\end{equation}
where $\tilde{a}_{in}^{\prime}$ is the optical input operator in
the laboratory frame and 
\begin{equation}
\omega_o=\omega_a-x_c g_{a0}-\frac{2G_a^2}{\omega_m}.
\end{equation}
Similarly the normal mode $B$ is microwave-like, 
\begin{equation}
\widetilde B_{\rm in}=\tilde b_{\rm in}^\prime e^{-i\omega_s t},\label{eq:SBint}
\end{equation}
with
\begin{equation}
\omega_s=\omega_b-x_c g_{b0}-\frac{2G_b^2}{\omega_m}.
\end{equation}

During the third stage of the detection sequence, $t>t_0+\tau$, the normal modes $A$ and $B$ switch their properties so that $\widetilde A_{\rm in}$ and $\widetilde B_{\rm in}$ will exchange their expressions in Eqs.~(\ref{eq:SAint}) and (\ref{eq:SBint}), but the form of Eq. (\ref{eq:SdVdt}) remains unchanged due to the symmetry. 

Finally, During the transduction stage, $t_0<t<t_0+\tau$, where we assume an adiabatic process without dissipation losses, the envelope field operators $\widetilde A$  and $\widetilde B$ only pick up phase factors due to the evolution,
\begin{equation}
\exp\left[-i\int_{t_0}^{t_0 + \tau}\frac{\partial\omega_{A(B)}}{\partial t}\,t\,dt\right].
\end{equation}
Since the period $\tau$ of this adiabatic transfer stage is much shorter than that of the other two stages and $\omega_{A(B)}$ changes slowly with time, we neglect the phase shift induced by this factor and assume $\hat V(t_0)=\hat V(t_0+\tau)$.  We can thus reexpress Eq.~(\ref{eq:SdVdt}) as
\begin{equation}
\partial_t\hat V=-\kappa\hat V+\sqrt{\kappa}\tilde a_{\rm in}^\prime e^{-i\omega_o t}+\sqrt{\kappa}\tilde b_{\rm in}^\prime e^{-i\omega_s t},\label{eq:Sdvab}
\end{equation}
which is now valid for the entire measurement sequence. By using the standard input-output relations for a two-sided cavity \cite{Walls94S},
\begin{eqnarray}
\tilde a_{\rm out}^\prime e^{-i\omega_o t} & = & \sqrt{\kappa}\hat V-\tilde a_{\rm in}^\prime e^{-i\omega_o t},\label{eq:Sdaoutp}\\
\tilde b_{\rm out}^\prime e^{-i\omega_s t} & = & \sqrt{\kappa}\hat V-\tilde b_{\rm in}^\prime e^{-i\omega_s t},
\end{eqnarray}
we can solve the dynamics in the frequency domain. By applying the Fourier transform to Eqs.~(\ref{eq:Sdvab}) and (\ref{eq:Sdaoutp}), we obtain
\begin{eqnarray}
i\omega\hat V(\omega) & = & -\kappa\hat V(\omega)+\sqrt{\kappa}\tilde a_{\rm in}^\prime(\omega+\omega_o) \nonumber \\
&+&\sqrt{\kappa}\tilde b_{\rm in}^\prime(\omega+\omega_s), \\
\tilde a_{\rm out}^\prime(\omega+\omega_o) & = & \sqrt{\kappa}\hat V(\omega)-\tilde a_{\rm in}^\prime(\omega+\omega_o),
\end{eqnarray}
which results into
\begin{equation}
\tilde a_{\rm out}^\prime(\omega+\omega_o)=\frac{\kappa\tilde b_{\rm in}^\prime(\omega+\omega_s)-i\omega\tilde a_{\rm in}^\prime(\omega+\omega_o)}{\kappa+i\omega}.
\end{equation}

\section{Numerical simulations}

\begin{figure*}
\includegraphics[width = \textwidth]{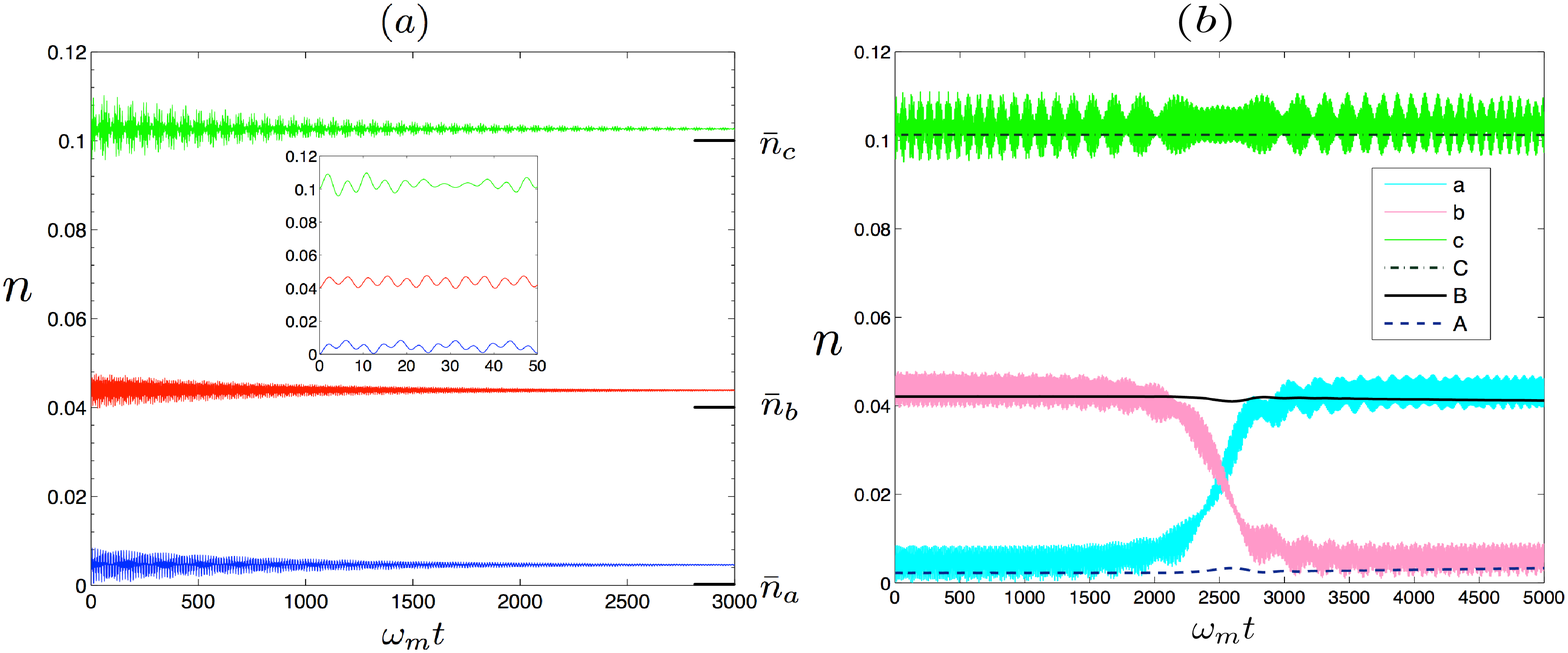}
\caption{(Color online) (a) Evolution of the populations of the optical mode $\hat a$ (blue), the microwave mode $\hat b $ (red), and the phonon mode $\hat c$ (green) starting from thermal states with mean populations $\bar n_a=0$, $\bar n_b=0.04$, and $\bar n_c=0.1$ (labelled by the tick marks on the right axis), respectively. The other parameters are $G_a=-G_b=0.05$, $\kappa_a=\kappa_b=\gamma=10^{-3}$, $\Delta_a=-0.5$, and $\Delta_b=-0.4$. The inset shows a detail of the short-time evolution.  (b) Evolutions of the populations of the bare modes and the normal modes during the transduction process. The optical detuning $\Delta_a$ is linearly varied from $-0.5$ to $-0.3$ with a constant microwave detuning $\Delta_b=-0.4$. Other parameters are  as in (a) but with decay rates $\kappa_a=\kappa_b=\gamma=10^{-5}$. All quantities are normalized to the phonon frequency $\omega_m$.\label{fig:Sconversion}}
\end{figure*}

Section II of the Supplemental Material showed that for the case of large phonon frequencies $\omega_m$, adiabatically eliminating the phonon mode $\hat c$  results in effective coupled Heisenberg-Langevin equations with squeezed noise for the optical mode $\hat a$ and the microwave mode $\hat b$, the two-mode model used in the main text, see also section IV of the Supplemental Material. To further validate this approximation we performed a numerical simulation of the master equation for the complete three-mode Hamiltonian
\begin{widetext}
\begin{eqnarray}
\frac{d\rho}{dt} & = & -\frac{i}{\hbar}[H,\rho]+\kappa_{a}(\bar n_a+1)\mathcal{L}[\hat a]\rho+\kappa_a\bar n_a\mathcal{L}[\hat a^\dagger]\rho\\
 & + & \kappa_b(\bar n_b+1)\mathcal{L}[\hat b]\rho+\kappa_b\bar n_b\mathcal{L}[\hat b^\dagger]\rho+\gamma(\bar n_c+1)\mathcal{L}[\hat c]\rho+\gamma\bar n_c\mathcal{L}[\hat c^\dagger]\rho,\nonumber 
\end{eqnarray}
where
\begin{equation}
H=\hbar\omega_m\hat c^\dagger\hat c-\hbar\Delta_a\hat a^\dagger\hat a-\hbar\Delta_b\hat b^\dagger\hat b+\hbar G_a(\hat a+\hat a^\dagger)(\hat c+\hat c^\dagger)+\hbar G_b(\hat b+\hat b^\dagger)(\hat c+\hat c^\dagger),
\end{equation}
\end{widetext}
 $\mathcal{L}$ represents the Lindblad superoperator of the form 
\begin{equation}
\mathcal{L}[\hat o]\rho=\hat o\rho\hat o^\dagger-\frac{1}{2}\hat o^\dagger\hat o\rho-\frac{1}{2}\rho\hat o^\dagger\hat{o}.
\end{equation}
and $(\kappa_a,\bar n_a)$, $(\kappa_b,\bar n_b)$, and $(\gamma,\bar n_c)$ are the decay rates and the mean thermal populations for the optical, microwave, and phonon modes, respectively.  We numerically evolved the master equation from an uncorrelated initial state $\rho(0)=\rho_a^{\rm th}\otimes\rho_b^{\rm th}\otimes\rho_c^{\rm th}$ with $\rho_{a,b,c}^{\rm th}$ the thermal state of the three modes, respectively.

As shown in Fig. \ref{fig:Sconversion}(a) the final steady population of the optical mode $\hat a$ and the microwave mode $\hat b$ are higher than their original mean thermal population $\bar n_a$ and $\bar n_b$ due to their interaction with the phonon mode $\hat c$. These increments, much smaller than the original thermal population, are found numerically to be of the order $0.004$ close to the estimate $0.003$ for the terms $m_{a,b}$ obtained from the adiabatic elimination in the Heisenberg-Langevin picture. The small mismatch, proportional to $G_{a,b}/\omega_m$ ,is due to the next-order contribution from the interaction between $\hat a$ and $\hat b$. The high-frequency oscillations of the populations during the evolution are determined by $\omega_m$ and $\Delta_{a,b}$, which are much faster than the decay rates $\kappa_{a.b}$, $\gamma$ that set the time scale to reach a steady state.

Fig. \ref{fig:Sconversion}(b) shows the transduction dynamics when the optical detuning is slowly switched from $\Delta_a<\Delta_b$ to $\Delta_a>\Delta_b$. The microwave mode $\hat b$ is initially in a thermal state with mean population $0.04$ while the optical mode $\hat a$ is initially in the vacuum state. The populations of the normal modes $\hat A$ and $\hat B$ represented by the blue dashed and the black thick curve, respectively, remain almost constant in the process, as expected for an adiabatic transformation. Meanwhile, their properties switch between optical-like and microwave-like around the resonant condition $\Delta_a=\Delta_b$.

The phonon mode $\hat c$ and the normal mode $\hat C$ appear to remain decoupled from the transduction process, validating the adiabatic elimination of the former.

\end{document}